\documentclass[10pt]{article}
\usepackage{amsmath}
\usepackage{graphicx}
\usepackage{color}
\usepackage{orcidlink}
\usepackage{hyperref}
\hypersetup{colorlinks=true, linkcolor=blue, citecolor=blue, urlcolor=blue}
\usepackage{caption}
\usepackage{subcaption}
\begin{document}
\baselineskip=14pt

\begin{center}
\LARGE {\bf Fermionic fields in a four-dimensional Bonnor-Melvin-Lambda space-time }
\end{center}

\vspace{0.3cm}

\begin{center}
    {\bf Faizuddin Ahmed\orcidlink{0000-0003-2196-9622}}\footnote{\bf faizuddinahmed15@gmail.com}\\
    \vspace{0.1cm}
    {\it Department of Physics, University of Science \& Technology Meghalaya, Ri-Bhoi, Meghalaya, 793101, India}\\
     \vspace{0.25cm}
    {\bf Nuray Candemir\orcidlink{0000-0002-6231-0499}}\footnote{\bf ncandemi@eskisehir.edu.tr (Corresp. author)}\\
     \vspace{0.1cm}
    {\it Department of Physics, Faculty of Science, Eskisehir Technical University, Eskisehir, Turkey}\\
    \vspace{0.25cm}
    {\bf Abdelmalek Bouzenada\orcidlink{0000-0002-3363-980X}}\footnote{\bf abdelmalekbouzenada@gmail.com  }\\
    \vspace{0.1cm}
    {\it  Laboratory of Theoretical and Applied Physics, Echahid Cheikh Larbi Tebessi University, Algeria}\\
    
\end{center}

\vspace{0.2cm}
\begin{abstract}
In this paper, we investigate how the gravitational field generated by a four-dimensional electrovacuum cosmological space-time influences the dynamics of fermionic fields governed by the Dirac equation, while also considering the effects of topology. We derive the radial wave equation corresponding to the relativistic Dirac equation and subsequently obtain analytical solutions for the energy levels and wave functions of the fermionic field within our chosen framework. Our analysis reveals that various parameters, including geometric topology, the cosmological constant, and quantum numbers, play significant roles in determining the eigenvalue solution of the quantum particles. Specifically, we demonstrate that the presence of the topological parameter disrupts the degeneracy of the energy spectrum.
\end{abstract}

\vspace{0.1cm}


\vspace{0.1cm}


\section{Introduction}
General relativity (GR) and quantum mechanics stand as the pillars of modern physics, offering profound insights into the behavior of matter and energy on both cosmic and subatomic scales. While general relativity provides a comprehensive framework for understanding the gravitational interactions between massive bodies and the curvature of spacetime, quantum mechanics revolutionizes our understanding of microscopic phenomena, revealing the probabilistic nature of particles and the wave-particle duality. The intersection of these two foundational theories gives rise to intriguing phenomena in theoretical physics, such as the elusive concept of quantum gravity and the exploration of physical phenomena within the context of spacetime singularities, including black holes and wormholes. This article explores the intricate interplay between general relativity and quantum mechanics, delving into the complexities of gravitational phenomena at the quantum level and their profound implications for our understanding of the universe \cite{r1,r2,r3,r4,r5,r6}.

The harmonic oscillator model serves as a cornerstone in understanding various phenomena in both classical and quantum mechanics. An intriguing question arises regarding how to appropriately describe this model within the framework of relativistic quantum mechanics. Initially, Ito {\it et al.} introduced a linear dependence in the radial coordinate into the Dirac equation using the minimal substitution \cite{M1}. They demonstrated that this modification leads to a harmonic oscillator with strong spin-orbit coupling. Subsequently, Moshinsky {\it et al.} reignited interest in this topic by employing the same minimal substitution procedure, referring to this system as the Dirac oscillator because it reduces to the usual quantum harmonic oscillator with a spin-orbit term in the non-relativistic limit \cite{M2}. Expanding on this concept, the full energy spectrum of the Dirac oscillator was derived in connection with supersymmetry \cite{M3}. A similar approach was taken in the context of the Klein-Gordon equation, resulting in the model known as the Klein-Gordon oscillator \cite{M4}.

In quantum mechanics, there is often a keen interest in examining how electromagnetic potentials impact the quantum dynamics of a given system. This investigation has been carried out within the framework of the Dirac oscillator. Researchers have delved into various aspects, including the study of the two-dimensional Dirac oscillator in the presence of a constant magnetic field \cite{M5} and the Aharonov-Bohm potential \cite{M6}. Furthermore, a path-integral formulation for the problem of a Dirac oscillator in the presence of a constant magnetic field has been presented in Ref. \cite{M7}. Coherent states for the (2 + 1)-dimensional Dirac oscillator coupled to an external field have also been constructed \cite{M8}. Reference \cite{M9} establishes a connection between the Dirac oscillator and the anti-Jaynes-Cummings model, elucidating explicit Landau levels. Additionally, the problem of a three-dimensional Dirac oscillator subjected to Aharonov-Bohm and magnetic flux potentials is addressed in Ref. \cite{M10}. The emergence of a relativistic quantum phase transition has been reported in the Dirac oscillator interacting with a magnetic field in both the usual oscillator \cite{M11} and the noncommutative oscillator \cite{M12}. Thermal properties of the Dirac oscillator in this context have also been explored \cite{M13, M14}. In Ref. \cite{M15}, the Dirac oscillator is analyzed in the presence of a magnetic field in an Aharonov-Bohm-Coulomb system. Another important aspect of studying quantum systems involves incorporating the influence of geometry on the physical properties of interest. In this regard, analyzing how curvature, for instance, affects a given system can be of interest \cite{M16}. Furthermore, exploring the quantum dynamics of a system immersed in a spacetime with a topological defect is also relevant \cite{M18}, as topological defects occur in various physical systems, spanning research areas such as cosmology \cite{M19} and condensed matter physics. The Dirac oscillator in backgrounds with topological defects has been investigated in several scenarios. For example, the energy spectrum \cite{M20} and coherent states \cite{M21} for the case of a cosmic string spacetime have been explored. The Dirac oscillator interacting with an Aharonov-Casher system in the presence of topological defects is studied in Ref. \cite{M22}. The cosmic string spacetime has been considered in the formulation of a generalized Dirac oscillator \cite{LFD} and also in the context of spin and pseudospin symmetries \cite{M24}.

The exploration of fermions in curved space has been extensively researched, as evidenced by the works of Parker \cite{MM22} and references \cite{MM23, MM24, MM25, MM26, M0, MM27, MM28, MM29, MM30, MMM3, PP2, PP3, PP4, PP5, PP6, PP7, EPJP, AOP, PP8, PP9}. This investigation extends into condensed matter and graphene domains, where fermions' behavior in curved space finds intriguing applications. Bakke \cite{MM31} offers interpretations of topological defects within the context of fermions in curved spacetime, presenting a Kaluza-Klein description of geometric phases in graphene. Stagmann and Szpak \cite{MM32} compare electronic transport in deformed graphene using both condensed matter and general relativistic approaches, the latter involving classical trajectories of relativistic point particles on curved surfaces. Iorio \cite{MM33} delves into the Hawking-Unruh phenomenon in graphene, while de Juan \cite{MM34} explores dislocations and twisting in graphene and related systems. Minář \cite{MM35} presents a perspective where a Dirac Hamiltonian in curved spacetime can be likened to a tight-binding Hamiltonian with non-unitary tunneling amplitudes. Experimental studies, such as those by Louko \cite{MM36}, examine the Unruh–DeWitt \cite{MM37} particle detector coupled to a scalar field in four-dimensional curved spacetime states. Boada \cite{MM38} utilizes cold atoms to create an artificial gravitational field, offering a platform to simulate fermions propagating in curved spacetime within this setup.

In this paper, we aim to investigate the dynamics of the fermionic fields within the context of a cosmological space-time described by the Bonnor-Melvin solution. In fact, we show that the eigenvalue solution of the fermionic field is influenced by the cosmological constant and is modified by the topological parameter. We organize this paper as follows: In Section 2, we derive the radial equation of the fermionic fields described by the Dirac equation in the considered space-time background. Then, we solve the radial equation using special functions, specifically the Nikiforov-Uvarov method, and obtain the energy levels and wave functions. In Section 3, we present our results.

\section{Fermionic Fields: The Dirac Equation }

In this part, we study the dynamics of spin-1/2 quantum particles described by the Dirac equation within the context of a cosmological constant background in four-dimensional space-time. We derive the radial wave equation and will obtain analytical solutions through special functions. Therefore, we begin this part by writing the line-element describing an electrovacuum cosmological space-time with the topological parameter $\alpha$ as given by \cite{MA, JV, MZ} (choosing system of units $c=1,\,\hbar=1$)
\begin{equation}
ds^2=-dt^2+dz^2+\frac{1}{2\,\Lambda}\,\Big(dr^2+\alpha^2\,\sin^2 r\,d\phi^{2}\Big),\label{a1}
\end{equation}
where the magnetic field strength becomes $H(r)=\sqrt{2}\,\alpha\,\sin r$ which is $\alpha$ times the original field strength obtained in \cite{MZ} and decreases by this amount. This space-time geometry has recently attracted much attention in studies of quantum mechanical problems (see, for examples \cite{FA1, FA2, FA3, FA4, FA5}). 

The covariant and contravariant metric tensor for the space-time (\ref{a1}) are given by
\begin{equation}
g_{\mu\nu}=\begin{pmatrix}
-1 & 0 & 0 & 0\\
0 & \frac{1}{2\,\Lambda} & 0 & 0\\
0 & 0 & \frac{\alpha^{2}\,\sin^2 r}{2\,\Lambda} & 0\\
0 & 0 & 0 & 1
\end{pmatrix},\quad\quad
g^{\mu\nu}=\begin{pmatrix}
-1 & 0 & 0 & 0\\
0 & 2\,\Lambda & 0 & 0\\
0 & 0 & \frac{2\,\Lambda}{\alpha^{2}\,\sin^{2} r} & 0\\
0 & 0 & 0 & 1
\end{pmatrix}.\label{a3}
\end{equation}
The determinant of the metric tensor $g_{\mu\nu}$ is given by
\begin{equation}
det\,(g_{\mu\nu})=g=-\frac{\alpha^2}{4\,\Lambda^2}\,\sin^2 r.\label{a4}
\end{equation}

For this diagonal space-time, we choose tetrad basis vector $e^{a}_{\mu}$ and its contravariant form $e^{\mu}_{a}$ given by
\begin{equation}
e_{\mu}^{a}=\begin{pmatrix} 1 & 0 & 0 & 0\\
0 & \frac{1}{\sqrt{2\,\Lambda}} & 0 & 0\\
0 & 0 & \frac{\alpha\,\sin r}{\sqrt{2\,\Lambda}} & 0\\
0 & 0 & 0 & 1
\end{pmatrix}, \quad\quad
e_{a}^{\mu}=\begin{pmatrix}
1 & 0 & 0 & 0\\
0 & \sqrt{2\,\Lambda} & 0 & 0\\
0 & 0 & \frac{\sqrt{2\,\Lambda}}{\alpha\,\sin r} & 0\\
0 & 0 & 0 & 1
\end{pmatrix}.\label{a5}
\end{equation}

The next step is a derivation of the Christoffel symbols for the considered space-time (\ref{a1}). It is define by
\begin{equation}
    \Gamma^{\lambda}_{\mu\nu}=\frac{1}{2}\,g^{\sigma\,\lambda}\,(g_{\mu\sigma,\nu}+g_{\nu\sigma,\mu}-g_{\mu\nu,\sigma}),\label{a6}
\end{equation}
where comma denotes ordinary derivative w. r. t argument. For the space-time (\ref{a1}), we obtain the following non-zero components given by
\begin{eqnarray}
    \Gamma^{r}_{\varphi\varphi}=-\alpha^2\,\sin r\,\cos r,\quad \Gamma^{\varphi}_{r\varphi}=\cot r.\label{a7} 
\end{eqnarray}

The spin connections $\omega_{\mu\,\,ab}$ can be determined using the Christoffel symbols $\Gamma^{\lambda}_{\mu\nu}$ with the following definition
\begin{equation}
\omega_{\mu\,\,b}^{a}=e_{\tau}^{a}\,e_{b}^{\nu}\,\Gamma_{\mu\nu}^{\tau}-e_{b}^{\tau}\,\partial_{\mu}\,e_{\tau}^{a}.\label{a8}
\end{equation}
For the space-time (\ref{a1}), the only non-zero component is given by
\begin{equation}
\omega_{\varphi\,a\,b}=\begin{pmatrix}0 & 0 & 0 & 0\\
0 & 0 & \frac{\alpha\,\cos r}{\sin r} & 0\\
0 & -\frac{\alpha\,\cos r}{\sin r} & 0 & 0\\
0 & 0 & 0 & 0
\end{pmatrix}.\label{a9}
\end{equation}
The spinorial affine connection $\varGamma_{\mu} ({\bf x})$ which is related with the spin connections $\omega_{\mu\,\,ab}$ defined by
\begin{equation}
    \varGamma_{\mu} ({\bf x})=-\frac{1}{2}\,\omega_{\mu\,ab}\,\Sigma^{ab},\quad\quad \Sigma^{ab}=\frac{1}{4}\,\left[\gamma^{a},\gamma^{b}\right].\label{a10}
\end{equation}

The corresponding Dirac equation in curved space with a non-minimal coupling of the background curvature $R$ with the field can be expressed by the following wave equation \cite{M17, MH, MMC}:
\begin{equation}
\Big[i\,\gamma^{\mu}\,\partial_{\mu}+i\,\gamma^{\mu}\,\varGamma_{\mu}-M\,c^2\Big]\,\Psi({\bf x})=0,\label{a11}
\end{equation}
where $M\,c^2$ is the rest mass energy with $M$ is the fermionic mass. In the subsequent discussion, we set $c=1$. Here, $\gamma^{\mu}$ are the gamma matrices in curved space related with the flat space gamma matrices given as follows:
\begin{equation}
\gamma^{\mu}({\bf x})=e_{a}^{\mu}\,\gamma^{a}.\label{a12}
\end{equation}
Here the flat gamma matrices are given by
\begin{equation}
    \gamma^0=\begin{pmatrix}
        {\bf I}_{2\times2} & {\bf 0}_{2\times2}\\
        {\bf 0}_{2\times2} & -{\bf I}_{2\times2}
    \end{pmatrix},\quad\quad 
    \gamma^i=\begin{pmatrix}
        {\bf 0}_{2\times2} & \sigma^i\\
        -\sigma^i & {\bf 0}_{2\times2}\\
    \end{pmatrix},\label{a13}
\end{equation}
where ${\bf I}$ is the unit matrix, ${\bf O}$ is the null matrix, and $\sigma^i$ are the Pauli matrices given by
\begin{equation}
    \sigma^1=\begin{pmatrix}
        0 & 1\\
        1 & 0
    \end{pmatrix},\quad
    \sigma^2=\begin{pmatrix}
        0 & -i\\
        i & 0
    \end{pmatrix},\quad
    \sigma^3=\begin{pmatrix}
        1 & 0\\
        0 & -1
    \end{pmatrix}.\label{a14}
\end{equation}

The gamma matrices $\gamma^{\mu}$ in curved space using Eq. (\ref{a12}) are given by
\begin{eqnarray}
    &&\gamma^t=e_{a}^{0}\,\gamma^{a}=\gamma^0 \Rightarrow \gamma^0\,\gamma^t=\mathcal{I},\nonumber\\
    &&\gamma^r=e_{a}^{1}\,\gamma^{a}=\sqrt{2\,\Lambda}\,\gamma^1 \Rightarrow \gamma^0\,\gamma^r=\sqrt{2\,\Lambda}\,\alpha^1,\nonumber\\
    &&\gamma^{\varphi}=e_{a}^{2}\,\gamma^{a}=\frac{\sqrt{2\,\Lambda}}{\alpha\,\sin r}\,\gamma^2 \Rightarrow \gamma^0\,\gamma^{\varphi}=\frac{\sqrt{2\,\Lambda}}{\alpha\,\sin r}\,\alpha^2,\nonumber\\ 
    &&\gamma^z=e_{a}^{3}\,\gamma^{a}=\gamma^3 \Rightarrow \gamma^0\,\gamma^{z}=\alpha^3,\label{a15}
\end{eqnarray}
where we have defined
\begin{equation}
   \alpha^i=\begin{pmatrix}
       0 & \sigma^i\\
       \sigma^i & 0
   \end{pmatrix}.\label{beta}    
\end{equation}

The spinorial affine connection $\varGamma_{\mu} ({\bf x})$ for the space-time (\ref{a1}) using Eq. (\ref{a10}) is given by
\begin{equation}
    \varGamma_{\varphi}=-\frac{1}{8}\,\omega_{\varphi\,ab}\,\left[\gamma^{a},\gamma^{b}\right]=-\frac{1}{8}\,\Big\{\omega_{\varphi\,12}\,\left[\gamma^{1},\gamma^{2}\right]+\omega_{\varphi\,21}\,\left[\gamma^{2},\gamma^{1}\right]  \Big\}=-\frac{1}{4}\,\omega_{\varphi\,12}\,\left[\gamma^{1},\gamma^{2}\right]\label{a16}
\end{equation}
since $\omega_{\varphi\,21}=-\omega_{\varphi\,12}$ and $\left[\gamma^{2},\gamma^{1}\right]=-\left[\gamma^{1},\gamma^{2}\right]$.

Using Eqs. (\ref{a9}) and (\ref{a13}), we obtain the non-zero spinorial affine connection given by
\begin{equation}
    \varGamma_{\varphi}=\frac{\alpha}{2}\,\frac{\cos r}{\sin r}\,i\,\sigma^3.\label{a17}
\end{equation}
Thus, we can find the following relation
\begin{equation}
    \gamma^{\mu}\,\varGamma_{\mu}=\gamma^{\varphi}\,\varGamma_{\varphi}=-\frac{i}{2}\,\frac{\cos r}{\sin r}\,\frac{\sqrt{2\,\Lambda}}{\sin r}\,\gamma^2\,\sigma^3 \Rightarrow \gamma^0\,\gamma^{\mu}\,\varGamma_{\mu}=\frac{i}{2}\,\frac{\cos r}{\sin r}\,\frac{\sqrt{2\,\Lambda}}{\sin r}\,\alpha^2\,\sigma^3.\label{a18}
\end{equation}

Thereby, the Dirac equation (\ref{a11}) using the curved space-time (\ref{a1}) can be expressed by the following linear differential equation
\begin{eqnarray}
\Bigg[i\,\partial_{t}+i\,\sqrt{2\,\Lambda}\,\alpha^1\,\partial_{r}+i\,\frac{\sqrt{2\,\Lambda}}{\alpha\,\sin r}\,\alpha^2\,\partial_{\varphi}+i\,\alpha^3\,\partial_{z}+\frac{1}{2}\,\frac{\cos r}{\sin r}\,\frac{\sqrt{2\,\Lambda}}{\sin r}\,\alpha^2\,\sigma^3-M\,\gamma^{0}\Bigg]\,\Psi=0.\label{a19}
\end{eqnarray}

To express the linear equation (\ref{a19}) explicitly, let us consider the following Dirac spinor field $\Psi (t, r, \varphi, z)$ given by
\begin{equation}
\Psi=e^{i\,\left[-\frac{E}{\hbar}\,t+\left(\ell+\frac{1}{2}\right)\,\varphi+k\,z\right]}\begin{pmatrix}
     \psi_1 (r)\\
     \psi_2 (r)
    \end{pmatrix},
\label{a20}
\end{equation}
where $\ell$ is the orbital quantum number, $E$ is the particle's energy, and $k$ is an arbitrary constant. In the subsequent discussion, we set $\hbar=1$.

Thereby, substituting Eq. (\ref{a20}) into the linear equation (\ref{a19}) and after simplification, we obtain the following set of differential equations for the spinor $\psi_i$ given  by
\begin{eqnarray}\label{a21}
&&\Big(E-M\Big)\psi_1=-i\,\sqrt{2\,\Lambda}\,\sigma^1\,\psi'_{2}+\Bigg[\frac{\left(\ell+\frac{1}{2}\right)\,\sqrt{2\,\Lambda}}{\alpha\,\sin r}\,\sigma^2+k\,\sigma^3+\frac{i}{2}\,\frac{\sqrt{2\,\Lambda} \cos r}{\sin^2 r}\,\sigma^1 \Bigg]\,\psi_2,\\
\label{a22}
&&\Big(E+M\Big)\,\psi_2=-i\,\sqrt{2\,\Lambda}\,\sigma^1\, \psi'_{1}+\Bigg[\frac{\left(\ell+\frac{1}{2}\right)\,\sqrt{2\,\Lambda}}{\alpha\, \sin r}\,\sigma^2+k\,\sigma^3+\frac{i}{2}\,\frac{\sqrt{2\,\Lambda}\,\cos r}{\sin^2 r}\, \sigma^1 \Bigg]\,\psi_1.
\end{eqnarray}

Decoupling the above set of equations (\ref{a21}) and (\ref{a22}), we obtain a second-order radial equation in terms of the first component of the Dirac spinor given by
\begin{eqnarray}\label{a23}
\psi''_{i}-\frac{\cos r}{\sin^2 r} \psi'_{i} +\Bigg(\frac{\cos^2 r}{2\,\sin^3 r}+\frac{\left(\ell+\frac{1}{2}\right)\,\sigma^3}{\alpha\,\sin^2 r}-\frac{\left(\ell+\frac{1}{2}\right)^2}{\alpha^2\,\sin^2 r}+ \frac{1}{2\sin^3 r}+\frac{\cos^2 r}{4\,\sin^4 r}\nonumber\\
+\frac{E^2-M^2-k^2}{2\,\Lambda}\Bigg)\psi_{i}=0.
\end{eqnarray}

Now, we convert this equation (\ref{a23}) into a standard form. Performing a transformation of the spinor $\psi_{i}(r)=e^{-\frac{\csc r}{2}}u(r)$, we obtain a differential equation in simpler form given by
\begin{equation}\label{a24}
u''(r)+\Big(\frac{\iota\,\sigma^3\cos r}{\sin^2 r}-\frac{\iota^2}{\sin^2 r}+\eta\Big)u(r)=0,
\end{equation}
where 
\begin{equation}
    \iota=\frac{\left|\ell+\frac{1}{2}\right|}{\alpha},\quad\quad \eta=\frac{E^2-M^2-k^2}{2\,\Lambda}.
\end{equation}

Note that $u(r)$ is an eigenfunction of $\sigma ^3$ whose eigenvalues $s=\pm\,1$, so we can write $\sigma ^3\,u=s\,u$, where $u=(u_{+}\,\,\,u_{-})^T$.

Choosing a change of variable $z=\cos r$, we can rewritten in Eq. (\ref{a24}) 
\begin{equation}\label{a25}
u''(z)-\frac{z}{1-z^{2}}\,u'(z)+\frac{1}{(1-z^2)^{2}}(-\beta_{1}z^{2}+\beta_{2}z-\beta_{3})\,u(z)=0,
\end{equation}
where we define the parameters
\begin{equation}\label{a26}
\beta_{1}=-\eta\quad,\quad \beta_{2}=\iota\, s \quad,\quad\beta_{3}=\iota^{2}-\eta.
\end{equation}

In quantum mechanical problems, various methods or techniques were employed in order to obtain exact and approximate analytical eigenvalue solutions. Among many of these, the Nikiforov-Uvarov (NU) method \cite{N1} is the most prominent method that has widely been used by several researchers (for examples, see \cite{NC,CA, KK1, KK2, KK3, KK4, KK5, KK6, KK7, KK8, KK9, KK10, KK11}). 

In this analysis, we employ this NU method in equation (\ref{a25}) and will obtain the eigenvalue solution. Now, if we apply the NU method by comparing Eq. (\ref{a25}) with the equation (\ref{eq.A.1}), we obtain 
\begin{equation}\label{a27}
\tilde{\tau}=-z\quad,\quad \sigma=1-z^{2} \quad,\quad\tilde{\sigma}(z)=\eta\, z^{2}+\iota\,s\, z-\iota^{2}+\eta.
\end{equation}
Substituting the above expression into Eq. (\ref{eq.A.8}) we obtain
\begin{equation}\label{a28}
\pi=-\frac{z}{2}\pm\sqrt{(\Upsilon-\kappa)z^{2}-\beta_{2}z+(\beta_{3}+\kappa)}\quad,\quad\Upsilon=\beta_{1}+\frac{1}{4}.
\end{equation}
According to the NU method, the values of $\kappa$ are found from the condition that the discriminant of the expression under square must be zero, so four possible values for $\pi(z)$ are founded
\begin{equation}\label{a29}
\pi(z)=\left\{\begin{array}{lll}
-\frac{z}{2}\pm\Bigg(\sqrt{\frac{\beta_{3}+\Upsilon+\sqrt{W}}{2}}z-\sqrt{\frac{\beta_{3}+\Upsilon-
\sqrt{W}}{2}}\Bigg)\ & \textrm{for}& \kappa= \frac{1}{2}\Bigg(\Upsilon-\beta_{3}-\sqrt{W}\Bigg)\\
-\frac{z}{2}\pm\Bigg(\sqrt{\frac{\beta_{3}+\Upsilon-\sqrt{W}}{2}} z+\sqrt{\frac{\beta_{3}+\Upsilon+\sqrt{W}}{2}}\Bigg) &\textrm{for}& \kappa= \frac{1}{2}\Bigg(\Upsilon-\beta_{3}+\sqrt{W}\Bigg),\\
\end{array}
\right.
\end{equation}
where $W=(\Upsilon+\beta_{3})^{2}-\beta_{2}^{2}$.

For the polynomial $\tau(z)$ in  (\ref{eq.A.4}) which has a negative derivative, we choose
\begin{equation}\label{a30}
\pi(z)=-\Bigg[\frac{1}{2}+\sqrt{\frac{\beta_{3}+\Upsilon+\sqrt{W}}{2}}\Bigg]z+\sqrt{\frac{\beta_{3}+\Upsilon-\sqrt{W}}{2}}.
\end{equation}
And
\begin{equation}\label{a31}
\kappa(z)=\frac{1}{2}\Bigg(\Upsilon-\beta_{3}-\sqrt{W}\Bigg).
\end{equation}
Thus, the polynomial $\tau(z)$ in Eq. (\ref{eq.A.4}) becomes
\begin{equation}\label{a32}
\tau(z)=-2z\Bigg[1+\sqrt{\frac{\beta_{3}+\Upsilon+\sqrt{W}}{2}}\Bigg]+2\sqrt{\frac{\beta_{3}+\Upsilon-\sqrt{W}}{2}}.
\end{equation}

Using Eqs. (\ref{eq.A.5}) and (\ref{eq.A.9}), the following expression for $\lambda$ and $\lambda_{n}$ are obtained respectively
\begin{equation}\label{a33}
\lambda=\frac{\Upsilon-\beta_{3}-\sqrt{W}}{2}-\frac{1}{2}-\sqrt{\frac{\Upsilon+\beta_{3}+\sqrt{W}}{2}}.
\end{equation}
\begin{equation}\label{a34}
\lambda_{n}=2\,n\Bigg(1+\sqrt{\frac{\Upsilon+\beta_{3}+\sqrt{W}}{2}}\Bigg)+n\,(n-1).
\end{equation}

After setting $\lambda=\lambda_{n}$, the energy eigenvalue equation is obtained
\begin{equation}\label{a35}
E^{\pm}_{n, \ell}=\pm\frac{\sqrt{(1+2\,\ell)^2\,\Lambda+\alpha^2\,\left[4\,k^2+3+Q+\sqrt{2}\,U+4\,n\,(2+2\,n+\sqrt{2} \,U)\,\Lambda+4\,M^2\right]}}{2\,\alpha},
\end{equation}
where 
\begin{equation}
    Q=\frac{\sqrt{[(1+2\,\ell)^2+\alpha^2]^2-4\,(1+2\,\ell)^2\,s^2\,\alpha^2}}{\alpha^2},\quad U=\sqrt{1+\frac{(1+2\,\ell)^2}{\alpha^2}+Q}.
\end{equation}

Equation (\ref{a35}) represents the relativistic energy eigenvalue of fermionic fields within the context of a magnetized universe generated by a cosmological constant and a geometric topology, known as the Bonnor-Melvin-Lambda curved space-time.

One can see that various parameters, such as the cosmological constant $\Lambda$, and the topology of the geometry characterized by the parameter $\alpha$, which is connected with the magnetic field strength $H(r)$, influence the energy levels. Moreover, the energy eigenvalue alters by the radial quantum number $n$, and the orbital quantum number $\ell$.

\begin{figure}[ht!]
\centering
\subfloat[$\ell=1,\,\alpha=1/2$]{\centering{}\includegraphics[scale=0.42]{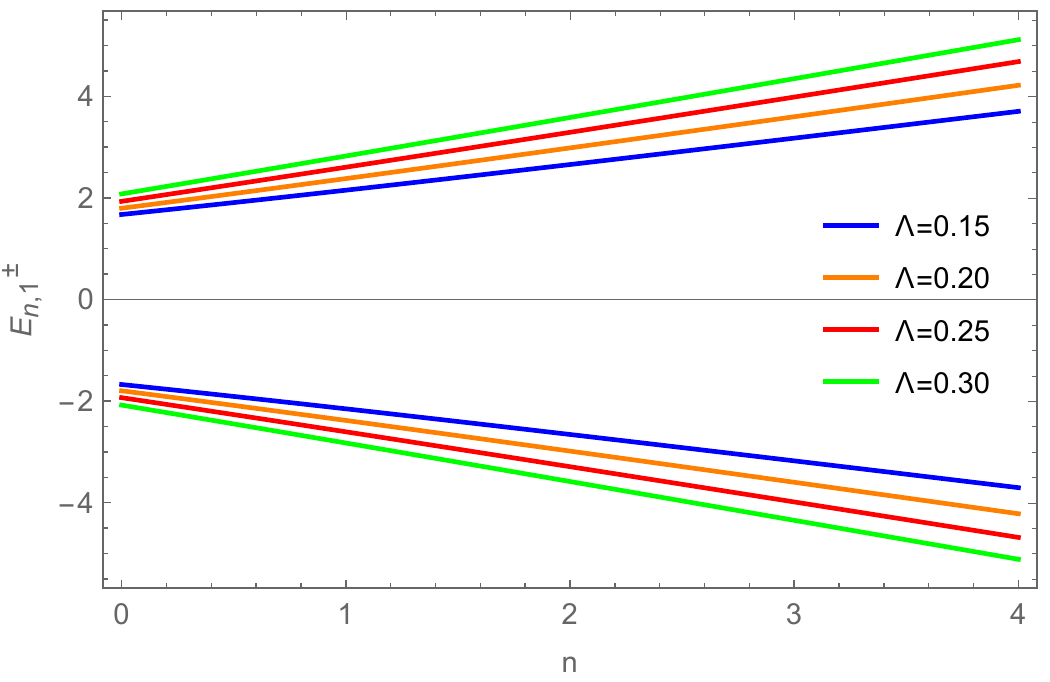}}\quad\quad
\subfloat[$\ell=1,\,\alpha=1/2$]{\centering{}\includegraphics[scale=0.42]{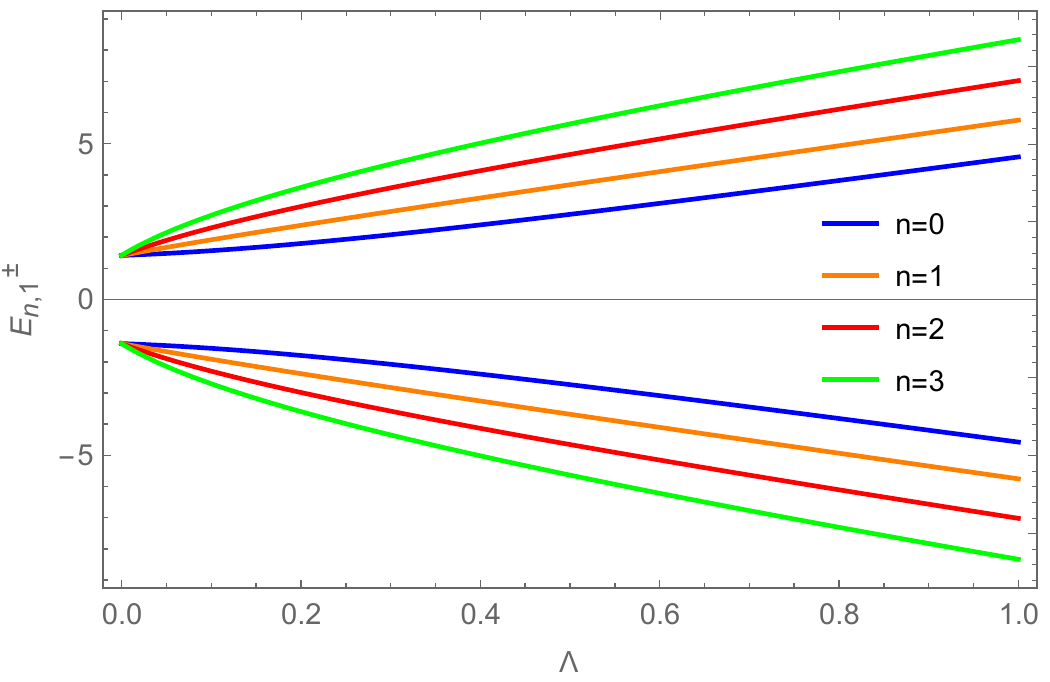}}
\hfill\\
\subfloat[$\ell=1,\,\Lambda=1/2$]{\centering{}\includegraphics[scale=0.42]{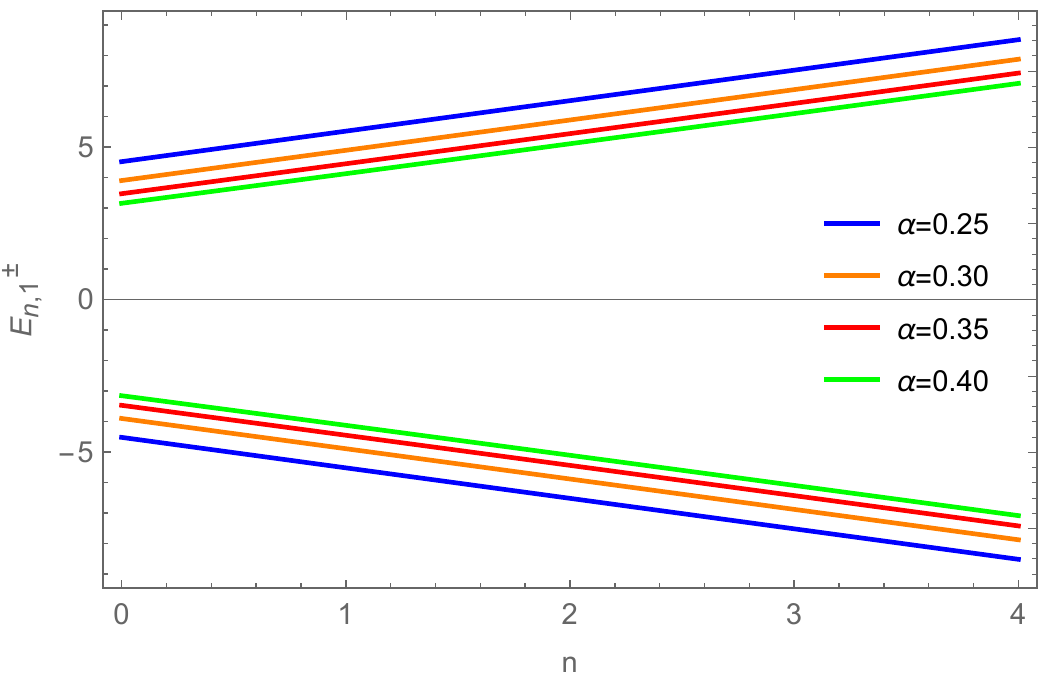}}\quad\quad
\subfloat[$\ell=1,\,n=1$]{\centering{}\includegraphics[scale=0.42]{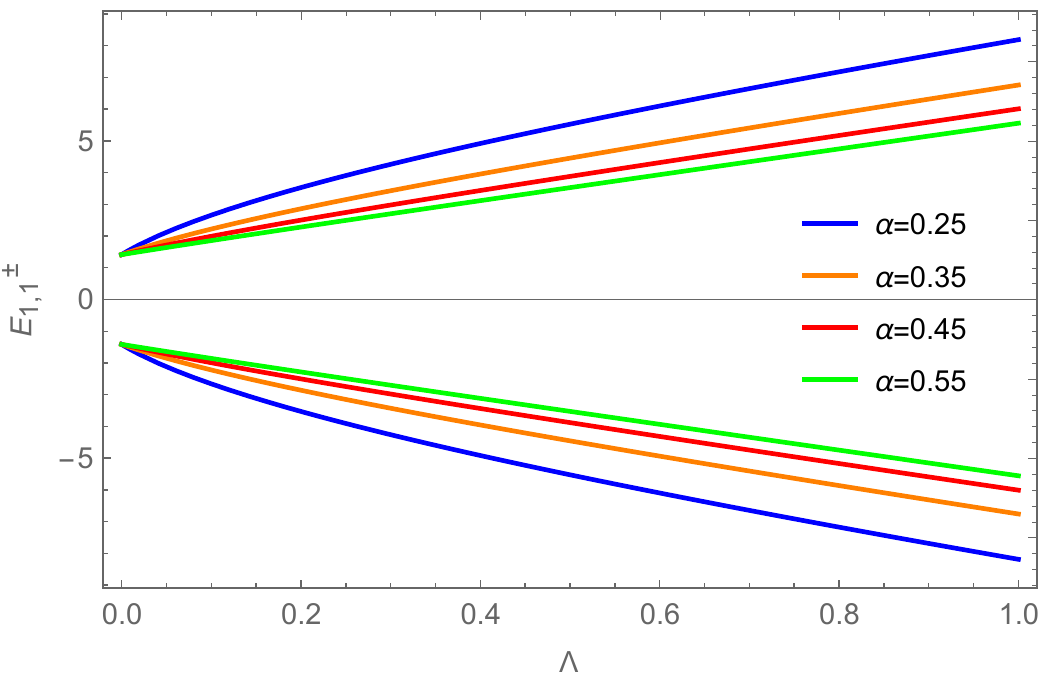}}
\hfill\\
\subfloat[$\ell=1$]{\centering{}\includegraphics[scale=0.42]{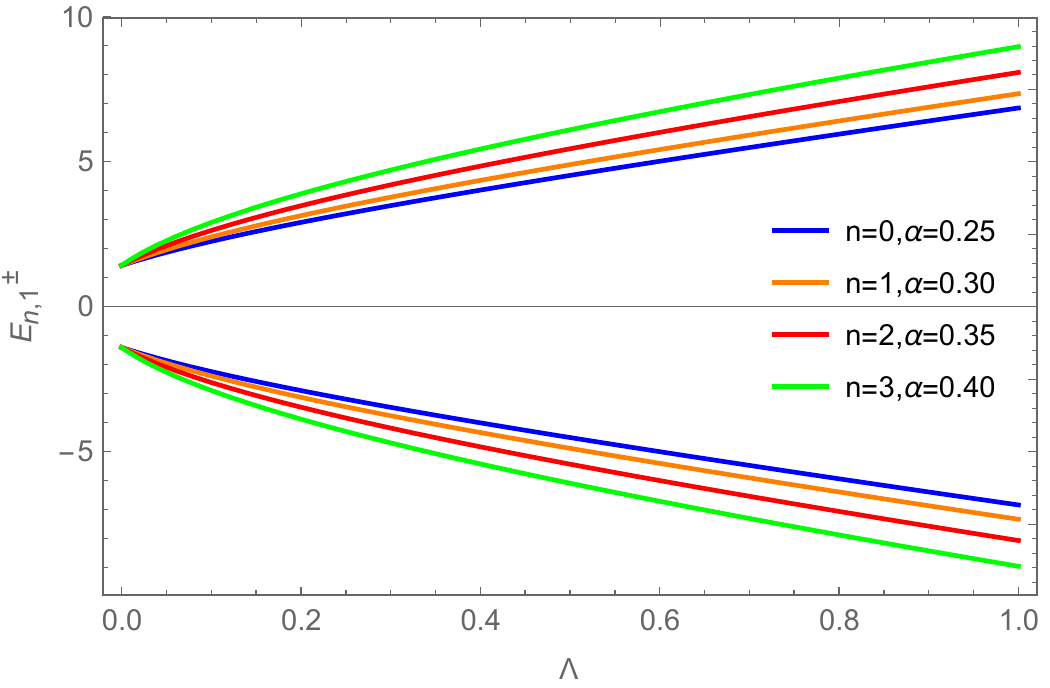}}\quad\quad
\subfloat[$\ell=1,\,n=1$]{\centering{}\includegraphics[scale=0.42]{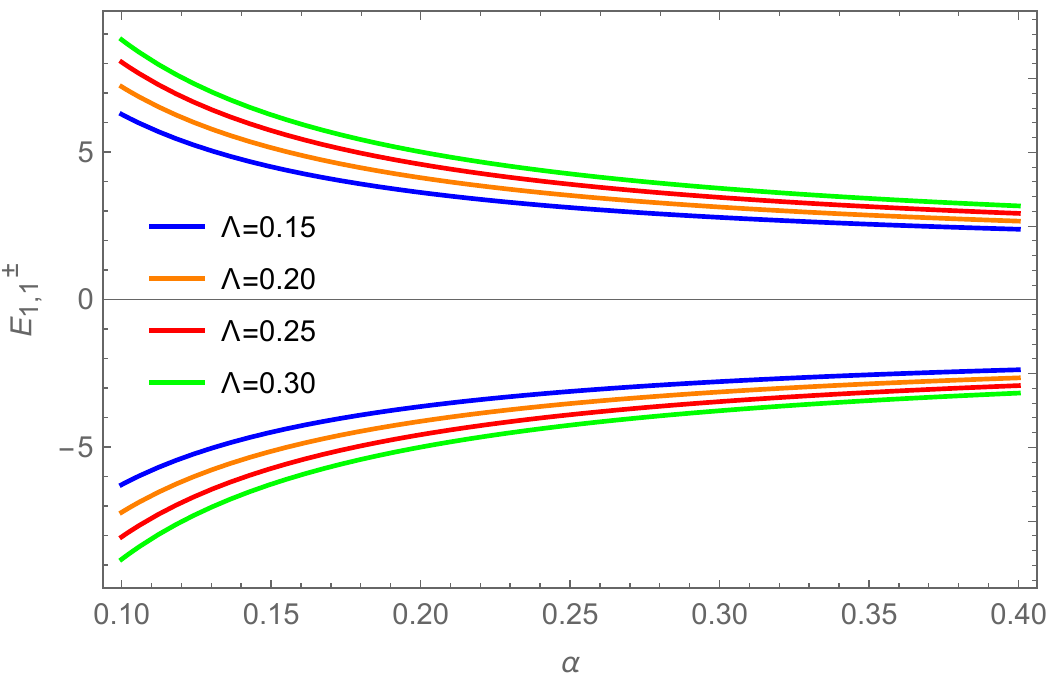}}
\centering{}\caption{The Energy levels of Eq. (\ref{a35}) with $n$ and $\Lambda$. Here, $M=1$ (energy unit), $k=1$ (energy unit), and $s=1/2$. We set the system of units, where $c=1=\hbar$.}
\label{Fig:1}
\end{figure}

We've generated Figure \ref{Fig:1}, illustrating the effects of various parameters-like the cosmological constant $\Lambda$, the topological parameter $\alpha$, and the quantum numbers $\{n, \ell\}$-on the energy spectrum of fermionic fields within the framework of Bonnor-Melvin cosmological space-time. In Figure 1(a)--1(b) and 1(e), it's evident that as the cosmological constant $\Lambda$ and the quantum number $n$ increase, the increasing energy levels shift upward. Conversely, in Figure 1(c)--1(d), these energy levels shift downward as the geometric topology parameter $\alpha$ increases. Additionally, in Figure 1(f), the decreasing energy levels shift upward as the cosmological constant $\Lambda$ increases.

Let us now find the corresponding wave function, using Eq. (\ref{a30}) and (\ref{a31}) with (\ref{eq.A.10}) and (\ref{eq.A.7}) we have
\begin{equation}\label{a36}
\phi(z)=(1-z)^{1/4(1+2K+2P)}(1+z)^{1/4(1+2K-2P)},
\end{equation}
\begin{equation}\label{a37}
\rho(z)=(1-z)^{(K-P)}(1-z)^{(K-P)},
\end{equation}
where $K=\sqrt{\frac{\beta_3+\Upsilon+\sqrt{W}}{2}}$ and $P=\sqrt{\frac{\beta_3+\Upsilon-\sqrt{W}}{2}}$. 

Then, inserting Eq. (\ref{a36}) into Eq.(\ref{eq.A.6}) , $y(z)$ is found
\begin{equation}\label{a38}
y_{n}(z)=B_{n}(1-z)^{(K-P)}(1+z)^{-(K+P)}\frac{d^{n}}{dz^{n}}\left[(1-z)^{(n+K-P)}(1+z)^{(n+K+P)}\right].
\end{equation}
and $B_{n}$ is the normalization constant. $y(z)$ are expressed in terms of Jacobi polynomial
\begin{equation}\label{a39}
y(z)=P_{n}^{(K+P,K-P)}(z).
\end{equation}

By using $u(z)=\phi(z)y(z)$, the solution of Eq. (\ref{a25})  is written
\begin{equation}\label{a41}
u(z)=(1-z)^{1/4(2K-2P+1)}(1+z)^{1/4(2K+2P+1)}P_{n}^{(K+P,K-P)}(z).
\end{equation}
Thus, the normalized radial wave function is given by the following expression 
\begin{equation}\label{a42}
\psi_{i} (r)=N e^{-\frac{\csc r}{2}}(1-\cos r)^{1/4(2K-2P+1)}(1+\cos r)^{1/4(2K+2P+1)}P_{n}^{(K+P,K-P)}(\cos r),
\end{equation}
where $N$ is a normalized constant.

\section{Conclusions}

Analyzing relativistic solutions for wave equations in various curved space-time backgrounds offers valuable insights into the behavior of quantum particles such as bosons and fermions in the presence of gravitational fields. Additionally, the presence of geometric topology also influences the quantum behavior of these particles. Researchers have investigated the dynamics of these quantum particles in various space-time solutions, including Schwarzschild, Kerr, Kerr-Newman, Gödel, and Gödel-type metrics, as well as cosmic string space-time with dislocations, disclinations, and point-like global monopole backgrounds.

In this research contribution, our aim was to study the dynamics of spin-1/2 quantum particles through the Dirac equation within the context of a cosmological space-time known as the Bonnor-Melvin solution in general relativity. We considered the effects of background curvature on the quantum motions of fermionic fields and analyzed the outcomes. We derived the radial equation of the Dirac wave equation describing the dynamics of fermionic fields within the context of Bonnor-Melvin-Lambda space-time, taking into account the effects of background curvature produced by the gravitational field. Using the Nikiforov-Uvarov method, we solved the decoupling equations for one spinor component and presented the relativistic energy levels and wave function of the fermionic fields. It has been observed that various factors such as the cosmological constant $\Lambda$ and the coupling constant $\xi$ influence the eigenvalues. Moreover, the eigenvalue solutions depend on the quantum numbers ${n, \ell}$ in addition to the spin factor $s$, changes values with change of them.

The Dirac fermion is a fundamental particle of great importance in both high-energy physics \cite{nn1,nn2} and condensed matter physics, including topological insulators \cite{nn3}, graphene, and related systems. Notably, in a one-dimensional lattice, the Dirac fermion exhibits key properties relevant to topological physics. Quantum simulation of Dirac fermions \cite{nn4} holds significant value, as these particles are ubiquitous in theoretical frameworks across physics. In recent years, topological phases have become one of the most captivating subjects in the field, with Dirac fermions playing a critical role \cite{nn5,nn6}.

Recently, experimental realizations of lattice topological models have been reported \cite{nn7}. Among the prominent experimental achievements, the realization of topological Thouless pumping \cite{nn8,nn9} and a ladder topological model in a synthetic dimensional optical lattice \cite{nn10} stand out. Although the Dirac fermion model, specifically on a lattice known as the Wilson-Dirac model \cite{nn11}, remains a theoretical framework without experimental quantum simulation, progress is being made. For instance, the authors of Ref. \cite{nn12} proposed a one-dimensional generalized Wilson-Dirac model (GWDM) as a promising quantum simulator.

Other experimental advances include the realization of Type-II Dirac fermions in the superconductor $\text{PdTe}_{2}$ \cite{nn13}, fractionalization waves in two-dimensional Dirac fermions \cite{nn14}, and the observation of massive Dirac fermions in a weak charge-ordering (CO) state in $\alpha-{(\text{BEDT-TTF})_2}\,\text{I}_3$ \cite{nn15}. We believe that the theoretical model of Dirac fermions presented in this study may serve as a useful foundation for future experimental work.

\section*{Appendix: The Nikiforov-Uvarov  (NU) Method }
\renewcommand{\theequation}{A.\arabic{equation}}

Nikiforov-Uvarov method is used to solve the second-order differential equation of hypergeometric type \cite{N1,NC}
\setcounter{equation}{0}
\begin{equation}
\label{eq.A.1}
\psi''(z)+\frac{\tilde{\tau}(z)}{\sigma(z)}\psi'(z)+\frac{\tilde{\sigma}(z)}{\sigma^2(z)}\psi(z)=0.
\end{equation}
Here, $\sigma(z)$ and $\tilde{\sigma}(z)$ are analytic polynomials, each at most of the second degree, and
$\tilde{\tau}(z)$ is a first degree polynomial. To find a particular solution of Eq. (A.1) by using the method of separation of variable, the $\psi(z)$ is written in terms of new functions
\begin{equation}
\label{eq.A.2}
\psi(z)=\phi(z)\,y(z).
\end{equation}
Thus, it reduced to a second-order differential equation
\begin{equation}
\label{eq.A.3}
\sigma(z)y''(z)+\tau(z)y'(z)+\lambda(z)y(z)=0,
\end{equation}
where
\begin{equation}
\label{eq.A.4}
\tau(z)=\tilde{\tau}(z)+2\pi(z)\quad \textrm{with}\quad\tau'(z)<0.
\end{equation}
And
\begin{equation}
\label{eq.A.5}
\lambda=\lambda_{n}=-n\tilde{\tau}(z)-\frac{n(n-1)}{2}\sigma''(z)\quad \textrm{with}\quad n=0,1,2..\,.
\end{equation}
Eq. (\ref{eq.A.3}) is the hypergeometric-type equation and $y(z)=y_{n}(z)$ is the hypergeometric-type function given by Rodrigues relation
\begin{equation}
\label{eq.A.6}
y_{n}(z)=\frac{B_{n}}{\rho(z)}\frac{d^n}{dz^n}[\sigma^n(z)\rho(z)],
\end{equation}
where $B_{n}$ is a normalization constant, and the weight function $\rho(z) $ must satisfy the condition
\begin{equation}
\label{eq.A.7}
[\sigma(z)\rho(z)]'= \tau(z)\rho(z),
\end{equation}
where the prime denotes the derivative concerning $z$. The term $\pi(z)$ is defined as
\begin{equation}
\label{eq.A.8}
\pi(z)=\frac{\sigma'(z)-\tilde{\tau}(z)}{2}\pm\sqrt{\Big(\frac{\sigma'(z)-\tilde{\tau}(z)}{2}\Big)^2-\tilde{\sigma}(z),
+\kappa\,\sigma(z)}
\end{equation}
and $\lambda$ parameter is expressed in terms of $\kappa$ and $\pi(z)$
\begin{equation}
\label{eq.A.9}
\lambda=\kappa+\pi'(z).
\end{equation}
The expression under the square root in Eq. (\ref{eq.A.8})  should constitute the square of  a first-degree polynomial since
$\pi(z)$ is a first degree polynomial. This condition holds true when the discriminant of the square root equals zero.
By utilizing this relation, $\kappa$ is determined and subsequently, $\pi(z)$ can be derived from the relevant $\kappa$ values. Eventually, comparing Eqs. (\ref{eq.A.5}) and (\ref{eq.A.9}), an eigenvalue equation values are obtained. A corresponding $\phi(z)$ is ascertained by computing the  logarithmic derivative
\begin{equation}
\label{eq.A.10}
\frac{\phi'(z)}{\phi(z)}=\frac{\pi(z)}{\sigma(z)}.
\end{equation}

\section*{Acknowledgements}

We sincerely acknowledged the anonymous referee for their valuable comments and helpful suggestions. F.A. acknowledged the Inter University Centre for Astronomy and Astrophysics (IUCAA), Pune, India for granting visiting associateship.







\end{document}